\documentclass[sigchi-a]{acmart}

\def\BibTeX{{\rm B\kern-.05em{\sc i\kern-.025em b}\kern-.08emT\kern-.1667em\lower.7ex\hbox{E}\kern-.125emX}}
    
\copyrightyear{2018}
\acmYear{2018}
\setcopyright{acmlicensed}
\acmConference[CHI Workshop '19]{New Directions for the IoT: Automate, Share, Build, and Care}{May 03--05, 2019}{Glasgow, UK}
\acmBooktitle{New Directions for the IoT: Automate, Share, Build, and Care, 2019, Glasgow, UK}
\acmPrice{15.00}
\acmDOI{10.1145/1122445.1122456}
\acmISBN{978-1-4503-9999-9/18/06}
\usepackage{hhline}
\usepackage{multirow}

\begin{document}

%
\title[On the Side Effects of Automation in IoT]{On the Side Effects of Automation in IoT:  Complacency and Comfort vs. Relapse and Distrust}

%
\author{Diego Casado-Mansilla}
\authornote{All authors contributed equally to this research.}
\email{dcasado@deusto.es}
\orcid{0000-0002-1070-7494}
\author{Pablo Garaizar}
\authornotemark[1]
\email{garaizar@deusto.es}
\affiliation{%
  \institution{University of Deusto}
  \streetaddress{Av. de las Universidades, 24}
  \city{Bilbao}
  \state{Spain}
  \postcode{48007}
}

\author{Anne M. Irizar}
\authornotemark[1]
\email{ane.irizar@deusto.es}
\author{Diego L\'opez de Ipi\~na}
\authornotemark[1]
\email{dipina@deusto.es}
\affiliation{%
  \institution{University of Deusto}
  \city{Bilbao}
  \state{Spain}
  \postcode{48007}
}

%
\renewcommand{\shortauthors}{Casado-Mansilla, et al.}

%
\begin{abstract}
Automation through IoT brings with it a whole new set of philosophical and ethical implications that we barely began to address. However, it is widely considered by many scholars as the panacea to overcome the majority of societal issues. The case of energy efficiency as an action for tackling climate change is not different: demand-response proposals or occupancy-driven energy management systems crowd the current research agenda on energy efficiency. However, there are still very few studies that have reported the effects of automation in the mid or long term beyond energy reduction (e.g. emotional feelings derived to interact with automation, complacency to the devices or perceived value of the automation throughout the time). In this workshop article, we report scientific evidence of a study conducted in ten workplaces during more than one year where we found that automating some electronic devices of common use (i.e. moving away or preventing subjects from the control of these devices) in favour of comfort and energy efficiency, is associated with a reduction of the users' confidence in science and technology as a mean to solve all environmental current problems and reduce the willingness of people to act in favor of the environment.

\end{abstract}

%

%

%
\maketitle

\section{Case-Study}

We carried out an experimental intervention of one year designed to test the effectiveness of the persuasive techniques in the mid and long-term because of some scholars raised their concern about the feasibility of persuasion to maintain the target behaviour throughout the time. In this study, we instrumented the electrical capsule-based coffee machines of ten different workplaces distributed between two big cities of Spain (Madrid and Bilbao) in order to record the energy consumption being drawn. The reasons why we selected these appliances were: \textit{1)} they are pretty common in work environments and are an element of shared use; \textit{2)} they consume large amounts of energy compared to other work appliances such as monitors or laptops. More than eighty people were recruited within the offices following a snowball procedure and their participation was voluntary (we raffled an energy monitoring system among participant who completed the whole study).

\begin{marginfigure}
\centering
\includegraphics[width=5cm, ,height=12cm]{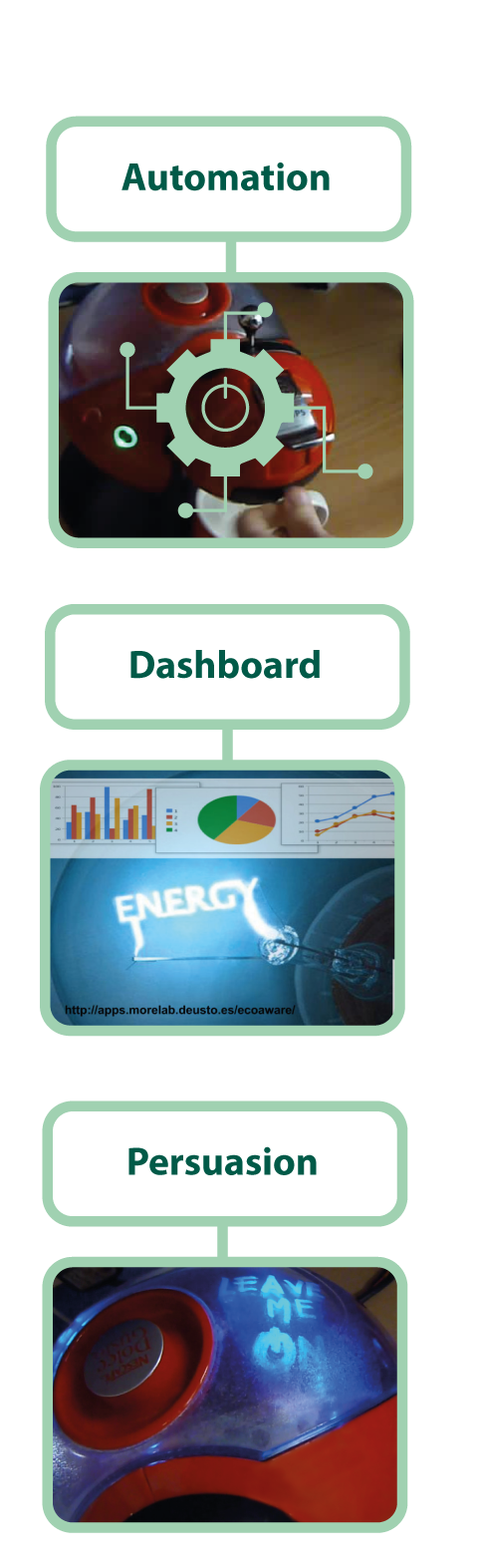}
\caption{The three different experimental treatments that were randomly assigned to the participating groups.}
\label{fig:treatments}
\end{marginfigure}

The study followed a between-group design approach. Thus, three different strategies to cope with energy inefficiency were tested among the participant groups (see Figure \ref{fig:treatments}). \textit{1) Persuasive feedback}: a combination of real-time ambient feedback and subtle visual hints to support the user's decision-making about when to switch off the appliance; \textit{2) Energy-dashboard }: participants were provided with a Web site to track their energy consumption associated with the appliance (i.e. self-monitoring and rational information through comparisons with historic energy data); and \textit{3) Automation }: the coffee makers were modified to autonomously switch the appliances off whenever they were not in use (i.e. the rationale behind automation was providing a sense of comfort to the users relieving them from the task of switching the appliance on and off after preparing a hot drink).

The random assignment of the experimental conditions among the ten workplaces remained as follows: Dashboard (3 workplaces), Automation (3 workplaces) and Persuasive feedback (4 workplaces). 

\subsection{Phases and Procedure}
The conducted study was divided into two phases: pre-pilot and post-piloting. During the former phase, the baseline of the energy wasted due to the misuse of capsule-based coffee machines (i.e. leaving the appliances in standby mode when not in use) was calculated. Furthermore, we asked participants ro respond an online questionnaire comprised of: 4 questions to obtain the socio-economic profile of each participant;  24 Likert-type questions pertaining to a questionnaire that evaluates the pro-environmental attitudes of the participants (Environmental Attitudes Inventory (EAI)~\cite{Milfont24EAI}; and 12 Likert-type questions belonging to the Pro-Environmental Readiness to Change Questionnaire (PE-RTC)~\cite{Tribble08PERB} that assesses the state or phases of change in which each of the participants was in relation to the intention to change their pro-environmental behavior: 'Pre-contemplation', 'Contemplation' or 'Action'.
At the end of the experiment (post-pilot), we calculated again the amount of energy being wasted in each coffee-maker and participants were requested to re-answer the 36  questions of the two online questionnaires: EAI and PE-RTC. Once the questionnaires' data was cleaned (e.g. remove outliers or uncompleted entries), we concluded that 81 participants appropriately responded in the pre-pilot phase and 48 participants did that accordingly at the end of the post-experimental phase. Hence, we removed the pre-pilot answers of 33 people in order to compare the results without introducing bias in the paired tests.
Finally, because of the energy findings and for triangulation purposes (i.e. a technique that facilitates validation of data through cross verification from two or more sources), we wanted to grasp detailed qualitative information from users to really understand the causes of the persuasive treatment being more effective than automation or dashboard and the users' reflections upon their assigned conditions. Because of the scope of this workshop, in the following, we only present the data resulting from the interaction with the Automated coffee-maker.

\section{Results}
At the end of the study, we found that the IoT-based Persuasive treatment helped to save most energy than the other two treatments reducing the energy waste by 44.53\%~\cite{Casado16Thesis}. The Automation treatment also helped to reduce energy waste by 14.19\%. Finally, the Dashboard approach did not lead to a reduction of energy waste remaining with a similar percentage as the beginning of the experiment.

\subsection{Statistical Analyses over EAI and PE-RTC Questionnaires}
In this section, the statistical analysis of the paired comparison between the pre and post-test responses for each of the participants associated with the automation-based coffee maker is presented. On the other hand, analysis of (co)variance is also provided using ANCOVA. The aim was to carry out an exploratory study to observe which factors and dimensions of pro-environmental attitudes (EAI) and intentions of pro-environmental change (PE-RTC) could be influenced by the fact of subjecting a group of users to the automation condition.

\subsubsection{Automation as a whole}
Sixteen people from three different working-groups interacted with the automated coffee machine. Comparing their responses between the pre-post piloting in the different constructs and scales of the questionnaires, we found statistical significance in the paired T-test on one of the scales of the EAI which evaluates the 'Trust in science and technology to solve all environmental problems': \textit{t (15) = 1.711, p = 0.0538}; the Effect Size (ES) using the Cohen's coefficient \textit{d = 0.427} with a confidence interval (CI) \textit{CI = [- 0.0916, 0.934]}. According to Cohen's power analysis criteria, this effect can be considered as medium~\cite{Cohen88Power}. Thus, the people who interacted with the automated coffee machine decreased their confidence in technology.

A sample of related questions in this sub-scale of the EAI inventory are:

\begin{itemize}
\item Science and technology will eventually solve our problems with pollution, overpopulation and
diminishing resources.
\item Modern science will solve our environmental problems.
\item We cannot keep counting on science and technology to solve our environmental problems.
\end{itemize}
Beyond this general analysis, we wanted to group working groups by socio-economic affinity by using hierarchical clustering. Having observed the similarities of the participant groups, we came out with three affinity groups, namely A, B and C. With this groups, we applied block design~\cite{Calinski00Block} in which the experimenter divides subjects into subgroups called blocks, such that the variability within blocks is less than the variability between blocks (the previous approach where all people assigned to automation condition were studied together). The results of the groupings, the conditions they were assigned to within groups are presented in Table \ref{tab:asig_experimental}.

\begin{margintable}
  \centering
  \resizebox{\linewidth}{!}{
    \begin{tabular}{
      >{\columncolor[HTML]{C0C0C0}}c |
      >{\columncolor[HTML]{EFEFEF}}l |
      >{\columncolor[HTML]{EFEFEF}}c |
      >{\columncolor[HTML]{EFEFEF}}c |
      >{\columncolor[HTML]{EFEFEF}}c |}
      \hhline{~----}
      \cellcolor[HTML]{FFFFFF} &  \textbf{Groups} & \textbf{C1} & \textbf{C2} & \textbf{C3} \\ 
      \cellcolor[HTML]{FFFFFF}& & \begin{scriptsize}Automation\end{scriptsize} & \begin{scriptsize}Dashboard\end{scriptsize} & \begin{scriptsize}Persuasion\end{scriptsize} \\  \hhline{~----}\hhline{~----}\hhline{~----}\hline
      \multicolumn{1}{|l|}{\cellcolor[HTML]{C0C0C0}}                                                                                             & Group 1 (8px)    & & & \textbf{X} \\ \cline{2-5}
      \multicolumn{1}{|l|}{\cellcolor[HTML]{C0C0C0}}                                                                                             & Group 2 (6px)        & \textbf{X} & & \\ \cline{2-5}
      \multicolumn{1}{|l|}{\cellcolor[HTML]{C0C0C0}}                                                                                             & Group 3 (3px) & & \textbf{X} & \\ \cline{2-5}
      \multicolumn{1}{|l|}{\multirow{-4}{*}{\cellcolor[HTML]{C0C0C0}\textbf{\begin{tabular}[c]{@{}c@{}}Grouping\\ Block A\end{tabular}}}}  & Group 4  (7px)      & & & \textbf{X} \\ \hline
      \multicolumn{1}{|l|}{\cellcolor[HTML]{C0C0C0}}                                                                                             & Group 5  (2px)        & &  & \textbf{X}\\ \cline{2-5}
      \multicolumn{1}{|l|}{\cellcolor[HTML]{C0C0C0}}                                                                                             & Group 6  (2px)      & \textbf{X} & & \\ \cline{2-5}
      \multicolumn{1}{|l|}{\multirow{-3}{*}{\cellcolor[HTML]{C0C0C0}\textbf{\begin{tabular}[c]{@{}c@{}}Grouping\\ Block B\end{tabular}}}} & Group 7 (4px)       & & \textbf{X} &  \\ \hline
      \multicolumn{1}{|l|}{\cellcolor[HTML]{C0C0C0}}                                                                                             & Group 8  (2px)     & & \textbf{X}&  \\ \cline{2-5}
      \multicolumn{1}{|l|}{\cellcolor[HTML]{C0C0C0}}                                                                                             & Group 9 (8px)      & \textbf{X} & & \\ \cline{2-5}
      \multicolumn{1}{|l|}{\multirow{-3}{*}{\cellcolor[HTML]{C0C0C0}\textbf{\begin{tabular}[c]{@{}c@{}}Grouping\\ Block C\end{tabular}}}}  & Group 10  (3px)       & &  & \textbf{X}\\ \hline
      
    \end{tabular}
    }
  \caption{Experimental assignment of each condition to the participant groups "blocked" by socio-economic affinity}
  \label{tab:asig_experimental}
\end{margintable}

\subsubsection{Automation within blocks of affinity}
Similarly, as the previous analysis studying the overall people assigned to one treatment, in Block A we found that users under automation condition showed lower reliability in science and technology: \textit{t(5) = 2.169, p = 0.0411}. ES was considered medium according to Cohen's criteria \textit{d = 0.547, CI = [- 0.893,1.987]}. Furthermore, applying ANCOVA over the three conditions, we found a difference in 'Confidence in science and technology': \textit{F$_{2, 20}$  = 2.872, p = 0.800} with an effect size $\eta_{p}^{2} =$  0.223 with  22.3\% of the variance explained by the automation experimental condition (posthoc Tukey was applied ) and 3.9\% due to the answers in the covariate: the Pre-experimental phase (the covariate is linearly related to the dependent variable and is not related to the condition).
Block B did not provide relevant differences in the studied constructs, however in Block C the subjects who interacted with the automatic coffee machines (automation) were found to decrease their active involvement in favor of the environment - the 'Action' state of change - (in this case, we applied Wilcoxon for non-parametric data):  $Z=$2.041, $p<$0.0312 with a large effect size measured in the Rosenthal coefficient $r =$ 0.510. Furthermore, in Block C we found a significant difference between the three experimental conditions in this 'Action' construct after applying ANCOVA analysis, being automation the condition which marked the difference:  $F_{2,9} = $11.264, $p = $0.0035 with an effect size $\eta_{p}^{2} =$0.714 which denotes that the automation experimental condition explains 71.4\% of the total variance, while that of the pre-test questionnaire (which is the covariate) just explains 7.7\% of the dependent variable.

\subsection{Analysis of Results}
The case of the automation condition aroused the greatest interest at a conclusive level. For the parametric tests carried out in the non-clustered approach, it was observed that subjects under this condition reduced their confidence in technology as a driver of pro-environmental change. When sampling noise was eliminated and the subjects were grouped in affinity blocks, it was observed that the confidence in the technology also decayed in Block A. Furthermore, the case of Block C was of relevance: the 8 people subjected to the automation condition reduced their state of Action (e.g. they were less active to act in favor of the environment). This result leads us to think that the condition has caused a significant reduction in pro-environmental attitudes and intentions.

Finally, taking into account the results from ANCOVA method, in Block A we observed that people subjected to automation presented at the end of the experiment the least perception of the use of technology as the main remedy for environmental problems. Besides, in group C the subjects under the automation condition significantly reduced their 'Action' state which entails that they seemed to be less active doing actions in favor of the environment.

\section{Discussion}
Keeping people away from the decision-making about turning the coffee machine on and off has not been free from critical and sympathized voices. On the one hand, the complacency induced by not having to think about how to act, knowing that the device operates by self-efficiently, was the main motivation of a large number of subjects interviewed at the beginning of the experiment. However, the same subjects showed certain rejection signs and frustration for not being able to manipulate the device at certain times (especially when the perception was that leaving the device switched on could cause energy waste). These feelings seem to be explanatory of the main finding of the attitudinal study: people under the automation condition showed less confidence in technology as the main pillar for climate change and environmental issue. Similarly, the people subjected to automation treatment were the ones among the three with the least confidence in science and technology.

In view of all the data presented, it seems that removing the subjects from the control of a simple electronic device of shared use in favor of comfort, reduced the users' confidence in the technology as a means to improve the overall environment in the future. Accordingly, it was observed that comfort due to automation may generate a rebound effect with respect to the passivity of acting in favor of the environment. As an example, in group C the subjects under the automation condition significantly reduced their 'Action' state. Applying this finding to the case of reducing the waste of energy in the workplace by applying automation; we claim that reducing the involvement of users in making simple decisions related to whether a machine should stay switched off or on may reduce their perception of energy expenditure, and therefore, may reduce their actions in favor of energy efficiency in such context.

\section{Conclusions and Future work}
In this article, we have witnessed an unexpected rebound effect caused by automation. According to the data presented, the fully automated management of processes focused on energy efficiency tends to generate a rebound effect causing, on the one hand, passivity to act in favor of the environment and, on the other, widespread distrust of science and technology. The results from the pre-post questionnaires and the interviews with participants of the study have shown that people may start feeling some sort of complacency by using an automated IoT device. However, when people do not fully understand certain processes of the automation and they lack control because of it, users might present initial states of reluctance to the automation. Taking this experiment as a baseline, there appears some  future lines of research. For instance,  to confirm this rebound effect and continue exploring why automation is contrary to some utilization factors of the pro-environmental attitudes. Furthermore, further research should be done through new experiments in groups with similar characteristics to Block A and C to assess if the results can be replicated and are specific for certain profiles.
%
\begin{acks}
This research is funded by HORIZON 2020 - (RIA)-696129-GREENSOUL. We also acknowledge the support of the Spanish government for  SentientThings under Grant No.: TIN2017-90042-R.
\end{acks}

%
\bibliographystyle{ACM-Reference-Format}
\bibliography{sample-base}

\end{document}